\documentclass[preprint,showpacs,preprintnumbers,amsmath,amssymb]{revtex4}

\usepackage{graphicx}
\usepackage{dcolumn}
\usepackage{bm}
\usepackage{color}

\usepackage{feynmf}
\usepackage{slashed}
\usepackage{enumerate}

\usepackage[utf8]{inputenc}

\begin{document}

\title{Probing the Light Sterile Neutrino Through the Heavy Charged Higgs Decay on the LHC}


\author{Yi-Lei Tang}
\thanks{tangyilei@kias.re.kr}
\affiliation{School of Physics, KIAS, 85 Hoegiro, Seoul 02455, Republic of Korea}

\date{\today}

\begin{abstract}
We show the 13 TeV proton-proton collider simulation in a $\nu$-two-Higgs-doublet-model ($\nu$-THDM). The heavy charged Higgs bosons are produced in pairs through the electroweak processes and decay to the light sterile neutrinos (lighter than the $W/Z$ boson masses). The light sterile neutrino further decays into a jet-like object with a muon in it. This helps us discriminate the signal from the backgrounds with the standard model jets.
\end{abstract}


\maketitle
\section{Introduction}

The see-saw mechanisms \cite{SeeSaw1, SeeSaw2, SeeSaw3, SeeSaw4, SeeSaw5} introduce some right-handed, or sterile neutrinos with extremely heavy Majorana masses ($\sim 10^{9\rm{-}12}$ GeV) to create the light neutrino masses ($\lesssim 0.1$ eV according to the oscillation data). However, such a sort of model is far beyond the ability for being probed on a collider. If we reduce the masses down to about 1 TeV, the Yukawa coupling among the sterile neutrinos, left-handed neutrinos and the Higgs doublet are predicted to be so small, that it is nearly impossible to produce a signal on a collider in reality. In the literature, there are some alternative models within which we can obtain a TeV-scale sterile neutrino as well as a relatively stronger connection with the standard model (SM) sectors. One group of such models utilizes the pseudo-Dirac sterile neutrinos (For examples, Ref.~\cite{Inverse1, Inverse2, Inverse3}) with a relatively larger Yukawa coupling $y$, yet the collider searching still involves hunting within an extremely small fraction of a $Z^{(*)}$/$W^{(*)}$-decay products (See \cite{DELPHISterileNeutrino, ATLASSterileNeutrino, CMSSterileNeutrino} for some experimental results. See \cite{ColliderInt1, ColliderInt2, LighterThanW, LightNHanTao, LeptonJets1, LeptonJets2, PseudoDiracSearch, WangKeChen}). Another approach on the collider is to associate the see-saw mechanisms with other new physics models \cite{NewAvenue1, NewAvenue2, NewAvenue3, NewAvenue4, NewAvenue5, NewAvenue6}. We can therefore look into the sterile neutrino with the aid of the other new physics particles.

In this paper, we consider a situation that the charged Higgs boson mainly decay to a sterile neutrino plus a charged lepton. This scenario can be found in some $\nu$-Two-Higgs-Doublet models ($\nu$THDM) \cite{vTHDM1, vTHDM2, vTHDM3, vTHDM4}. The collider phenomenology has also been studied in the literature, too \cite{vTHDM_Collider1, vTHDM_Collider2}. However, as we know, an interesting parameter space $m_{N} \ll m_{H^{\pm}}$, and $m_{N} \lesssim m_{Z/W}$ remains to be a gap in careful study. In this area, each largely-boosted sterile neutrinos decay into a single jet-like collimated object so that the usual method for the separated objects loses effectiveness (For some neutrino jet works, see Ref.~\cite{NeutrinoJet1, NeutrinoJet2}). Sometimes, if the $m_N < m_{W/Z}$, the three-body suppression on the sterile neutrino's decay width will reveal a secondary vertex for us to discriminate the signal (as being addressed in the Ref.~\cite{vTHDM_Collider1}). However, the decay length depends on the model parameters and is possible to be well below the $1$ cm scale. Therefore, in our discussions, we neglect all the secondary vertex informations and treat the signal/SM backgrounds in the usual way. Unlike most of the SM-jets,  the sterile neutrino jets usually contain leptons. This feature has been applied in, e.g.,  the Ref.~\cite{LeptonJets1, LeptonJets2} to search for the lepton jets decayed from the sterile neutrinos. However, we mainly consider the hadronical decay and other new physics sector participates the processes in this paper. We also apply this feature to eliminate the background efficiently. 

\section{Model Description}

Here, we rely on a standard $\nu$-THDM model. The THDM with the $\Phi_i \rightarrow (-1)^{i-1} \Phi_i$ $Z_2$ symmetry including the softly-breaking terms is characterized by the effective potential \cite{THDMInt}
\begin{eqnarray}
  V&=& m_1^2 \Phi_1^{\dagger} \Phi_1 + m_2^2 \Phi_2^{\dagger} \Phi_2 - m_{12}^2 (\Phi_1^{\dagger} \Phi_2 + \Phi_2^{\dagger} \Phi_1) + \frac{\lambda_1}{2} (\Phi_1^{\dagger} \Phi_1)^2 + \frac{\lambda_2}{2} (\Phi_2^{\dagger} \Phi_2)^2 \nonumber \\
  &+& \lambda_3 (\Phi_1^{\dagger} \Phi_1)(\Phi_2^{\dagger} \Phi_2) + \lambda_4 (\Phi_1^{\dagger} \Phi_2)(\Phi_2^{\dagger} \Phi_1) + \frac{\lambda_5}{2} \left[ (\Phi_1^{\dagger} \Phi_2)^2 + (\Phi_2^{\dagger} \Phi_1)^2 \right],
\end{eqnarray}
where $\Phi_{1,2}$ are the two Higgs doublets with the hypercharge $Y=\frac{1}{2}$, and $\lambda_{1-7}$ are the coupling constants, $m_{1, 2, 12}^2$ are the mass parameters. The $\nu$-THDM is based upon the Type-I THDM in which all the SM particles $Q_L$, $u_R$, $d_R$, $L_L$, $e_R$ couple with the $\Phi_2$ field
\begin{eqnarray}
\mathcal{L}_{\text{Yukawa}}^{\text{SM}} = -Y_{u i j} \overline{Q}_{L i} \tilde{\Phi}_2 u_{R j} - Y_{d i j} \overline{Q}_{L i} \Phi_2 d_{R j} - Y_{l i j} \overline{L}_{L i} \Phi_2 l_{R j} + \text{h.c.}.
\end{eqnarray}
The sterile neutrino together with the left-handed lepton doublets couple with the $\Phi_1$. In this paper, without loss of generality, we consider only one sterile neutrino $N$. Therefore the corresponding Lagrangian is given by
\begin{eqnarray}
\mathcal{L}_{\text{Yukawa}}^{\nu} =  - m_N \overline{N} N - ( Y_{i} \overline{L}_{L i} \tilde{\Phi}_1 N + \text{h.c.}),
\end{eqnarray}
where $m_N$ is the mass for the sterile neutrino and $Y_i$, $i=1,2,3$ are the Yukawa coupling constants corresponding to the $e$, $\mu$, $\tau$ lepton doublets.

In the following discussions, we do not need to care about the details of the electroweak symmetry breaking, nor do we need to discuss the neutral Higgs bosons. After the electroweak symmetry breaking, we acquires a coupling
\begin{eqnarray}
\mathcal{L} \supset -Y_i \sin \beta H^+ \overline{l}_i P_R N + \text{h.c.}, \label{ChargedYukawa}
\end{eqnarray}
where $\tan \beta = \frac{v_2}{v_1}$ is the ratio of the $\Phi_{1,2}$ vacuum expectation values, and $H^{\pm}$ is the charged Higgs with the mass of $m_{H^{\pm}}$. In the large $\tan \beta \gg 1$ case, the (\ref{ChargedYukawa}) becomes the most significant coupling for the $H^{\pm}$ to decay, and therefore $H^{\pm} \rightarrow l^{\pm} N$ will become the dominant decay channel when $m_N < m_{H^{\pm}}$.

On a proton-proton collider, the $H^+ H^-$ pairs can be produced via either s-channel $\gamma^*$, $Z^*$, or off-shell Higgs particles $h^*$, $H^*$. The off-shell Higgs particle channel is usually negligible due to the rather small light quark-Higgs coupling constants, except in some special cases when $\lambda_3$ or $\lambda_4$ is very large. Therefore in this paper, we only consider the $q \overline{q} \rightarrow Z^*/\gamma^* \rightarrow H^+ H^-$ processes. Another important thing that we should mention is the decay length of the sterile neutrino $N$. In fact, the smallness of the left-handed neutrino masses constrain the $Y_i \cos \beta < 10^{-6}$, which in turn amplify the decay length up above $\gtrsim 1$ m. This will destroy most of our results discussed in this paper. However, the model we apply in this paper is only a simplified model. In reality, this can be quite different in the case of the existence of the other sterile neutrino singlets. For example, in the Ref.~\cite{MyPaper}, the appearance of the pseudo-Dirac sterile neutrinos will allow a much larger $Y_i \cos \beta \sim 10^{-3}$ in a naturally smaller $\tan \beta$ case.  Most of the results that our paper will remain undisturbed in such cases. Furthermore, the current bounds from various experiments on the mixing parameters $|U_{iN}|^2 \approx \left( \frac{Y_i \cos \beta v}{m_N} \right)^2$ are no less than $10^{-6}$. In this paper, we only care about the parameter space near these bounds, just as many theoretical and experimental works in the literature. Therefore we do not consider the possibility of the secondary vertex cases.

\section{Simulation Details and Results}

\begin{figure}
\includegraphics[width=3in]{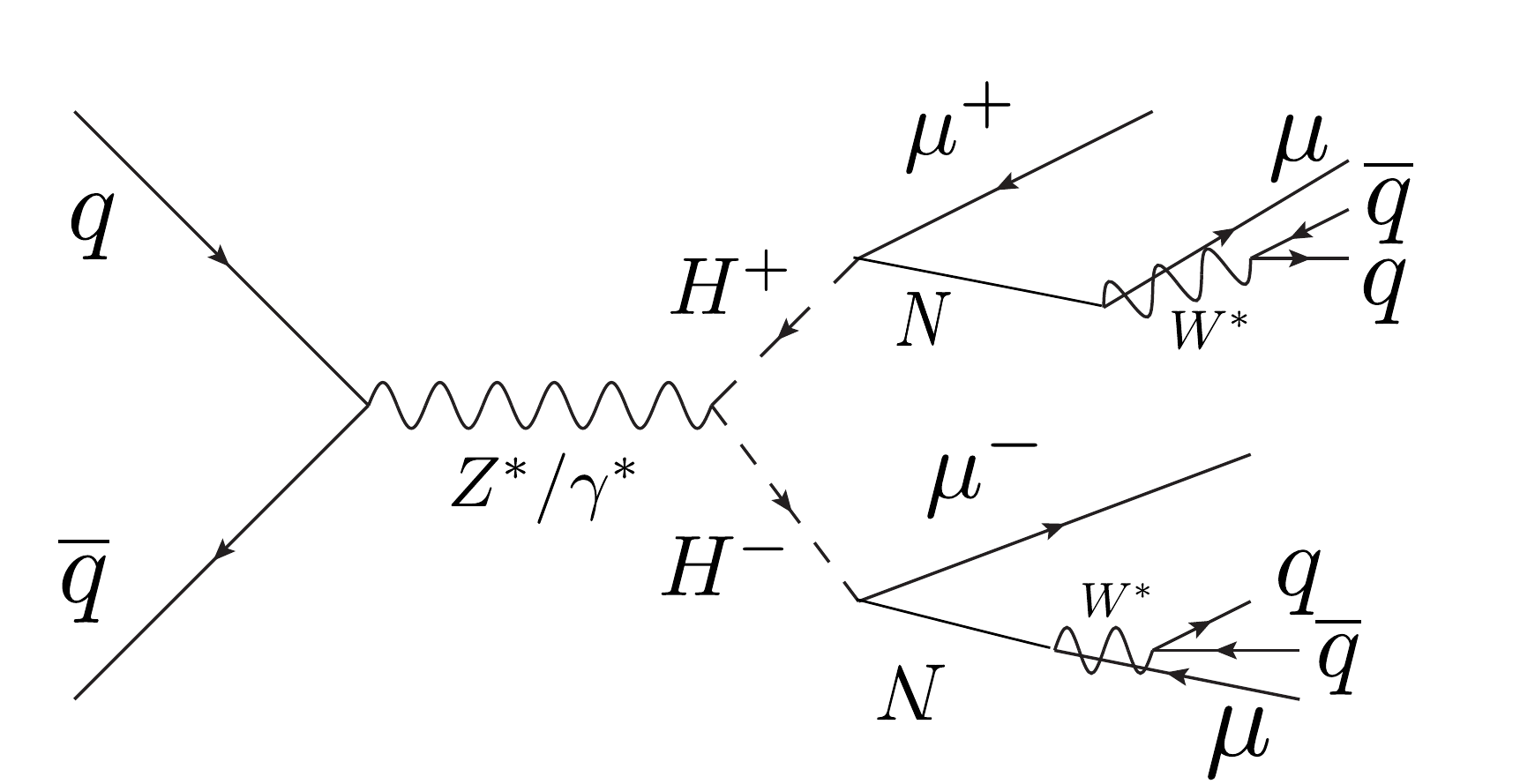}
\caption{$p p \rightarrow Z^*/\gamma^*  H^+ H^-$, with the $H^{\pm} \rightarrow \mu^{\pm} N$, $N \rightarrow \mu^{\pm} W^{\mp *} \rightarrow \mu^{\pm} q \overline{q}$ decay chains. Since $m_N \ll m_{H^{\pm}}$, the decay products for $N$ will be largely boosted into two jet-like objects.} \label{MyProcess}
\end{figure}

In this paper, we concentrate on the $p p \rightarrow Z^*/\gamma^*  \rightarrow H^+ H^-$, with the $H^{\pm} \rightarrow \mu^{\pm} N$, $N \rightarrow \mu^{\pm} W^{\mp *} \rightarrow \mu^{\pm} q \overline{q}$ decay chains as shown in the Fig.~\ref{MyProcess}. We have chosen the hadronic decay channel of the sterile neutrino because of its largest branching ratio and the convenience to reconstruct the $H^{\pm}$ masses. The muon appeared in the decay products (clustered inside a jet together with the other elements) can help us tag the jets decayed from the sterile neutrinos. 

The muons appeared in the decay products can also be replaced by electrons or taus. Here we have chosen the muon channels for the collider's distinctive ability to identify a muon, especially for a muon inside a jet. As for the electron cases, we can estimate the corresponding SM-background from the result in this paper due to the leptonic universality in the SM. However, one should be aware of the more difficulties in identifying an electron inside a jet than a muon in reality when he wants to infer our result to the other lepton cases.

The missing energy is predicted to be small in our expected signal. The reducible backgrounds with neutrinos ought to be eliminated by applying the transverse missing energy (MET) cuts. However, the pile-up effects on future high luminosity colliders might seriously smear the MET distributions in both the signal and the background events, making it difficult to separate them. There are three main sources of the reducible backgrounds with neutrinos: from the $\tau^+ \tau^-$'s leptonic decay, from the gauge bosons $W^+ W^- \rightarrow l^+ \nu l^- \overline{\nu}$ decay and $Z Z \rightarrow \nu \overline{\nu} \mu^+ \mu^-$ decay associated with $b \overline{b}$, and the $t \overline{t} \rightarrow b \overline{b} W^+ W^-$ decay. The $\tau^+ \tau^-$'s leptonic decay is suppressed by the relatively small branching ratio, and these less energetic muons have much less chances to pass the kinematic cuts. gauge boson decay channels are further reduced by the higher order of the coupling constants. The only significant reducible channel is the $p p \rightarrow t \overline{t} \rightarrow b \overline{b} \mu^+ \mu^- \nu \overline{\nu}$. Therefore, in this paper, we only consider the $t \overline{t}$ channel among all the reducible backgrounds. Delphes with its CMS card shows that the MET distribution of the signal sample is also rather significant even in the case without pileup. Such MET might due to the mismeasurement of a jet momentum and usually the direction of the missing energy is parallel to one of the jets, therefore $\Delta \phi_{\text{MET-}j}$ can be a ideal parameter to cut the reducible background. However, in this paper, the jets are from the generator-level particles for a preliminary study, and the future techniques and performances of the missing energy measurements and the influence of pile-up effects are unknown. Therefore, in this paper, we do not apply the MET cuts during the event selection processes and show our results in two situations: $p p \rightarrow t \overline{t}$ is not considered (naming ``no-$t \overline{t}$''), or it is considered (naming ``with-$t \overline{t}$''). These indicate the two extreme situations that the reducible backgrounds can be fully cut out or not in the future.

We generate our events through the FeynRules+MadGraph5\underline{~}aMC+PYTHIA8+Delphes \cite{FeynRules, MadGraph, PYTHIA8, Delphes}. The informations of the isolated muons come from the Delphes results. However, we cluster the generator-level particles (with the isolated muons deducted) by ourselves with the FastJet. The event selection criteria are given by:
\begin{itemize}
\item two isolated muons with at least two $p_T \geq 50$ GeV and $|\eta|<=5.0$ jets appear. The isolated muons are selected according to the CMS card settings in the Delphes, and the jets are clustered by the anti-$k_T$ method with the parameter $R=0.9$ by all the $|\eta|<=5.0$ particles. The two isolated muons do not participate the jet clustering processes.
\item The invariant mass between the two muons should be outside the $Z$-boson mass window $[75, 105]$ GeV. The invariant mass between the two $p_T$ leading jets should be outside the mass window $[60, 130]$ GeV.
\item Group the two muons and two jets into two pairs. Each pair contains one muon and one jet. The absolute value of the invariant mass difference should be the smallest among all grouping possibilities (2 possibilities for each event). The invariant mass of each pair should be within the $[m_{H^\pm}-50\rm{ GeV}, m_{H^\pm}+50\rm{ GeV}]$ range.
\end{itemize}

The above are the pure kinematic criteria. We suffix the event samples which have only passed these selections by `a `-KIN''. For example, ``SIG-KIN'' means the signal events which have only passed these selections, while ``BKG-KIN'' indicates the corresponding background events.

We then tag the jets as the decay product of a sterile neutrino by examining each constituent within the jet. If there is a muon carrying more than $30\%$ of the total jet energy, then this jet is tagged as a ``N-jet''. The events containing at least one tagged N-jet is suffixed by ``-1N-jet'', and the ones with two tagged N-jet is suffixed by ``-2N-jet''. We should also note that we have assumed that 100\% of the muons inside a jet can be perfectly identified by the detector, with no mis-identification rate of the other particles. (For some muon identification discussions on the current CMS detector, see Ref~.\cite{CMS-PAS-PFT-10-003, CMS-muon, CMS-muon-earlier}. One can find the identification and misidentification rates there if he needs a careful simulation. Ref.~\cite{CMS-muon-earlier} had also mentioned to identify a muon inside a b-jet.)

Leptons rarely appear in a SM jet, although there are some semi-leptonic decay channels from a short-lived hadron, especially for B-mesons. Both the b-jets and the other jets have a non-ignorable chance to fake a sterile-neutrino jet, however the probability is so small that it is rather difficult to generate a background ``-2N-jet'' sample. Nevertheless, we can divide the tagged jet number by the total jet number to estimate the ``mistagging rate'' $R_{\mu}^{j,b}$ for each $jjll$ or $bbll$ samples, the ``$j$'' or ``$b$'' indicate the $udcsg$-jets or the $b$-jets respectively. Finally we can calculate the $R_{\mu}^{j 2}$, $R_{\mu}^{b 2}$, or $R_{\mu}^{j} R_{\mu}^{b}$ times ``BKG-KIN'' cross sections to estimate the two-sterile-neutrino-jet background (``BKG-2N-jet'' cross sections).

\begin{table}
\begin{tabular}{|c|c|c|c|c|c|c|c|}
\hline
$m_{H^{\pm}}/$GeV & 100 & 125 & 150 & 175 & 200 & 225 & 250 \\
\hline
$\sigma_{\rm{BKG-KIN}}/$fb & 1.43e+03 & 1.11e+03 & 730 & 470 & 301 & 193 & 129 \\
\hline
$\sigma_{\rm{BKG-1N-jet}}/$fb & 5.31 & 3.8 & 2.58 & 1.43 & 0.686 & 0.588 & 0.407 \\
\hline
$\sigma_{\rm{BKG-2N-jet}}/$fb & 0.0106 & 0.00858 & 0.00562 & 0.0036 & 0.00239 & 0.00155 & 0.00103 \\
\hline
$m_{H^{\pm}}/$GeV & 275 & 300 & 325 & 350 & 375 & 400 & \\
\hline
$\sigma_{\rm{BKG-KIN}}/$fb & 86.3 & 58.2 & 39.3 & 27.9 & 19.1 & 13.3 & \\
\hline
$\sigma_{\rm{BKG-1N-jet}}/$fb & 0.252 & 0.169 & 0.106 & 0.0845 & 0.0609 & 0.0467 & \\
\hline
$\sigma_{\rm{BKG-2N-jet}}/$fb & 0.000617 & 0.000413 & 0.000293 & 0.000213 & 0.000161 & 0.00011 & \\
\hline
\end{tabular}
\caption{Background cross sections corresponding to each mass window around $m_{H^{\pm}}$ and different sterile-neutrino-jet number criteria. $p p \rightarrow t \overline{t} \rightarrow \mu^+ \mu^- b \overline{b} \nu \overline{\nu}$ contributions are not included.} \label{Background}
\end{table}

\begin{table}
\begin{tabular}{|c|c|c|c|c|c|c|c|}
\hline
$m_{H^{\pm}}/$GeV & 100 & 125 & 150 & 175 & 200 & 225 & 250 \\
\hline
$\sigma_{\rm{BKG-KIN}}/$fb & 2.06e+03 & 1.76e+03 & 1.2e+03 & 724 & 427 & 271 & 181 \\
\hline
$\sigma_{\rm{BKG-1N-jet}}/$fb & 31.4 & 27.8 & 17.3 & 7.74 & 3.56 & 2.51 & 1.57 \\
\hline
$\sigma_{\rm{BKG-2N-jet}}/$fb & 0.183 & 0.186 & 0.134 & 0.0733 & 0.0371 & 0.0228 & 0.0152 \\
\hline
$m_{H^{\pm}}/$GeV & 275 & 300 & 325 & 350 & 375 & 400 & \\
\hline
$\sigma_{\rm{BKG-KIN}}/$fb & 121 & 81.2 & 54.3 & 37.5 & 25.6 & 17.9 & \\
\hline
$\sigma_{\rm{BKG-1N-jet}}/$fb & 1 & 0.612 & 0.481 & 0.306 & 0.214 & 0.132 & \\
\hline
$\sigma_{\rm{BKG-2N-jet}}/$fb & 0.0101 & 0.00673 & 0.0044 & 0.00283 & 0.00195 & 0.00138 & \\
\hline
\end{tabular}
\caption{The same with the \ref{Background}. However, $p p \rightarrow t \overline{t} \rightarrow \mu^+ \mu^- b \overline{b} \nu \overline{\nu}$ contributions are included.} \label{Background_tt}
\end{table}

We have classified the background channels according to the final states as well as the mediator resonances and generate the corresponding samples separately. Besides the $p p \rightarrow t \overline{t}$ contributions, the main background channels are $p p \rightarrow \mu^+ \mu^- j j$, $p p \rightarrow \mu^+ \mu^- j b$ and $p p \rightarrow \mu^+ \mu^- b b$ with no $W/Z$ resonances, The $W/Z$ resonance channels such as $p p \rightarrow ZZ \rightarrow \mu^+ \mu^- j j$ are also considered and analysed separately. The background cross sections corresponding to different sterile-neutrino-jet number criteria (``-1N-jet'' or ``-2N-jet'') on a 13 TeV proton-proton collider are listed in Tab.~\ref{Background} and \ref{Background_tt}. In the Tab.~\ref{Background}, $p p \rightarrow t \overline{t} \rightarrow \mu^+ \mu^- b \overline{b} \nu \overline{\nu}$ channels are not included, while in the Tab.~\ref{Background_tt}, these contributions are included.

\begin{table}
\begin{tabular}{|c|c|c|c|c|c|c|c|}
\hline
$m_{H^{\pm}}/$GeV & 100 & 125 & 150 & 175 & 200 & 225 & 250 \\
\hline
$\sigma_{pp\rightarrow H^+ H^-}/$fb & 204 & 91.2 & 47 & 26.5 & 15.9 & 10.1 & 6.65 \\
\hline
$R_{\rm{SIG-KIN}}$ & 0.154 & 0.159 & 0.16 & 0.212 & 0.214 & 0.208 & 0.206 \\
\hline
$R_{\rm{SIG-1N-jet}}$ & 0.102 & 0.106 & 0.106 & 0.142 & 0.144 & 0.139 & 0.136 \\
\hline
$R_{\rm{SIG-2N-jet}}$ & 0.0275 & 0.0284 & 0.0284 & 0.0389 & 0.0379 & 0.0373 & 0.0364 \\
\hline
$m_{H^{\pm}}/$GeV & 275 & 300 & 325 & 350 & 375 & 400 & \\
\hline
$\sigma_{pp\rightarrow H^+ H^-}/$fb & 4.53 & 3.17 & 2.26 & 1.65 & 1.22 & 0.916 & \\
\hline
$R_{\rm{SIG-KIN}}$ & 0.253 & 0.254 & 0.251 & 0.252 & 0.247 & 0.296 & \\
\hline
$R_{\rm{SIG-1N-jet}}$ & 0.174 & 0.174 & 0.17 & 0.166 & 0.163 & 0.204 & \\
\hline
$R_{\rm{SIG-2N-jet}}$ & 0.0498 & 0.0477 & 0.0474 & 0.0461 & 0.0431 & 0.0576 & \\
\hline
\end{tabular}
\caption{The total cross section of $p p \rightarrow H^+ H^-$ on a 13 TeV proton-proton collider and the cut efficiencies of the signal corresponding to each mass window on $m_{H^{\pm}}$ and different sterile-neutrino-jet number criteria. The sterile neutrino mass $m_N$ is fixed at 10 GeV.} \label{SignalData}
\end{table}

Besides the electroweak $p p \rightarrow H^+ H^-$ cross sections $\sigma_{p p \rightarrow H^+ H^-}$, the signal cross sections also depend on the branching ratios of the $H^{\pm} \rightarrow \mu^{\pm} N$  and $N \rightarrow \mu^{\pm} q \overline{q}$ decay channels. All these factors can be characterized by one single parameter $\epsilon$, defined by $\sigma_{\mu-\text{ channels}}=\epsilon \sigma_{p p \rightarrow H^+ H^-}$, or $\epsilon = \left( \text{Br}_{H^{\pm} \rightarrow \mu^{\pm} N} \cdot \text{Br}_{N \rightarrow \mu^{\pm} q \overline{q}} \right)^2$. We have calculated the total $p p \rightarrow H^+ H^-$ cross sections as well as the cut efficiencies $R_{\text{SIG-X}}$ for each $m_{H^{\pm}} \in [100, 400]$ GeV in a 25 GeV interval and $m_{N} \in [10, \frac{m_{H^{\pm}}}{10\text{GeV}}+5]$ GeV in a 2.5 GeV interval. We do not discuss the case of $m_N < 10 \text{ GeV}$ because the PYTHIA8 might not be able to give us a reliable showering and hadronization result in this case. Part of the direct result is listed in the Tab.~\ref{SignalData}. In the Tab.~\ref{SignalData}, we fix the $m_N$ at 10 GeV for the limited size of the table.

\begin{figure}
\includegraphics[width=3.2in]{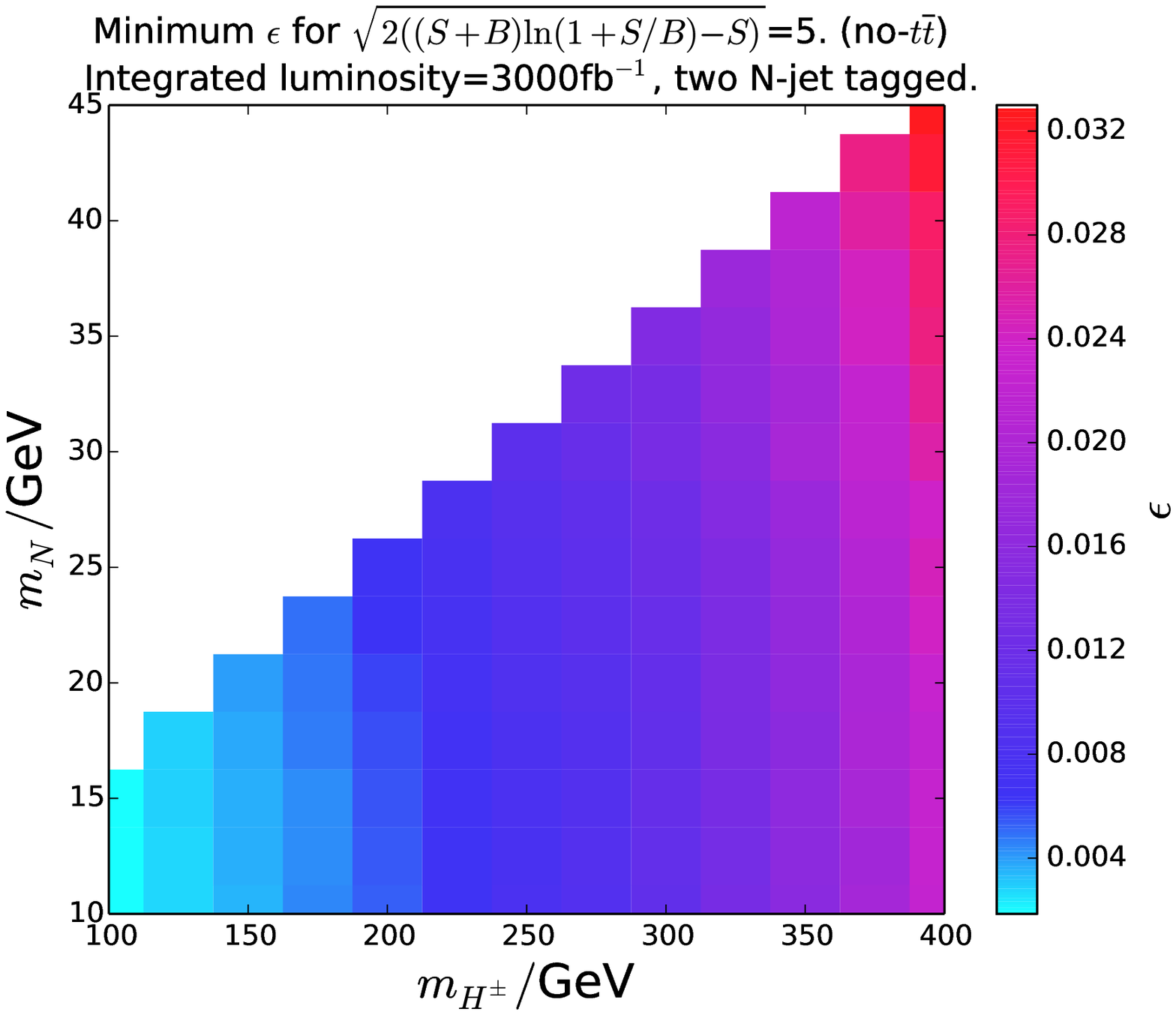}
\includegraphics[width=3.2in]{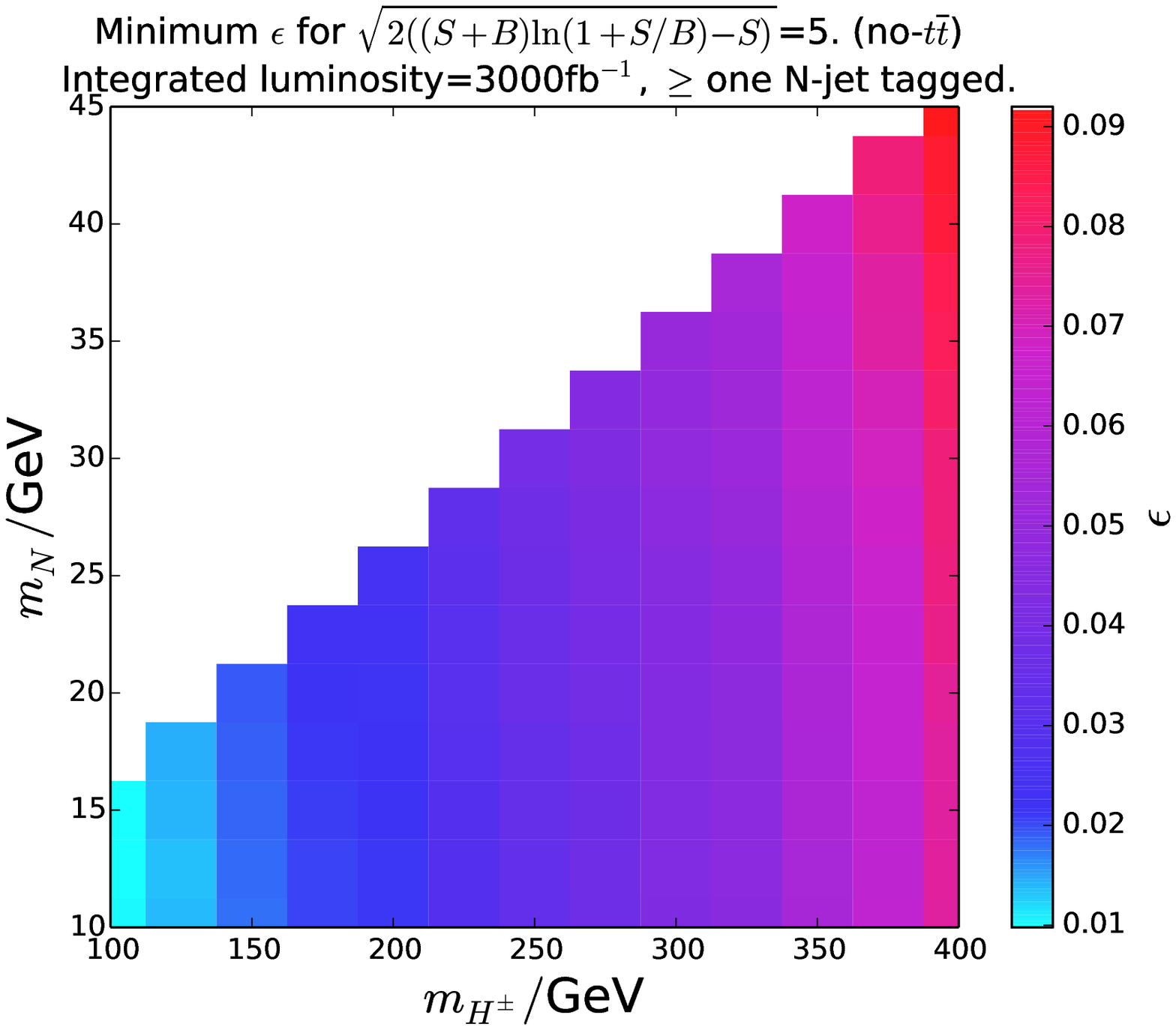}
\caption{Minimum $\epsilon$ for $\sqrt{2 ((S+B) \ln(1+S/B)-S)} = 5$. The integrated luminosity is set 3 ab$^{-1}$ on a 13 TeV proton-proton collider. $p p \rightarrow t \overline{t} \rightarrow \mu^+ \mu^- b \overline{b} \nu \overline{\nu}$ contributions are not included.} \label{MinimumEpsilon_3000}
\end{figure}

\begin{figure}
\includegraphics[width=3.2in]{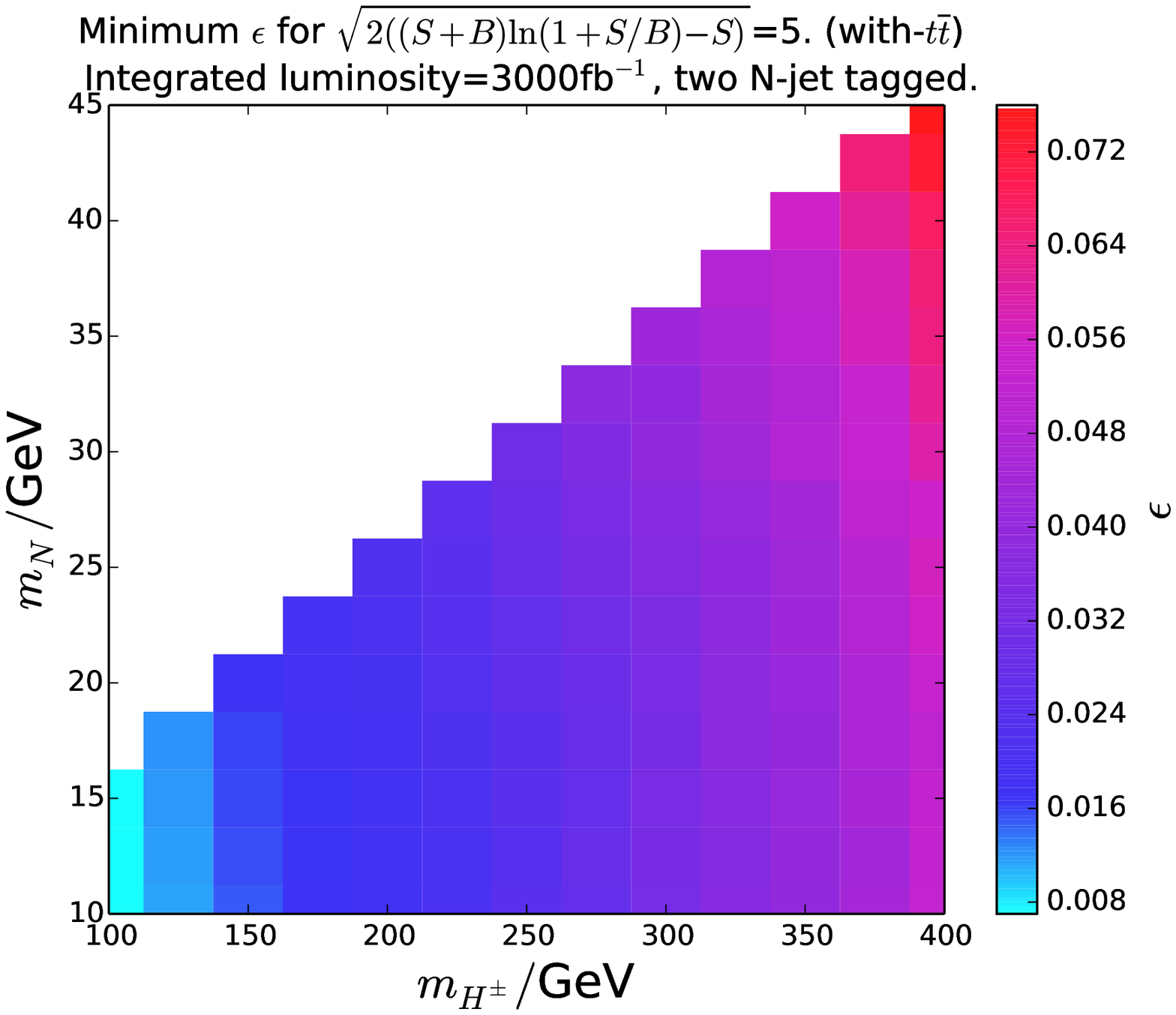}
\includegraphics[width=3.2in]{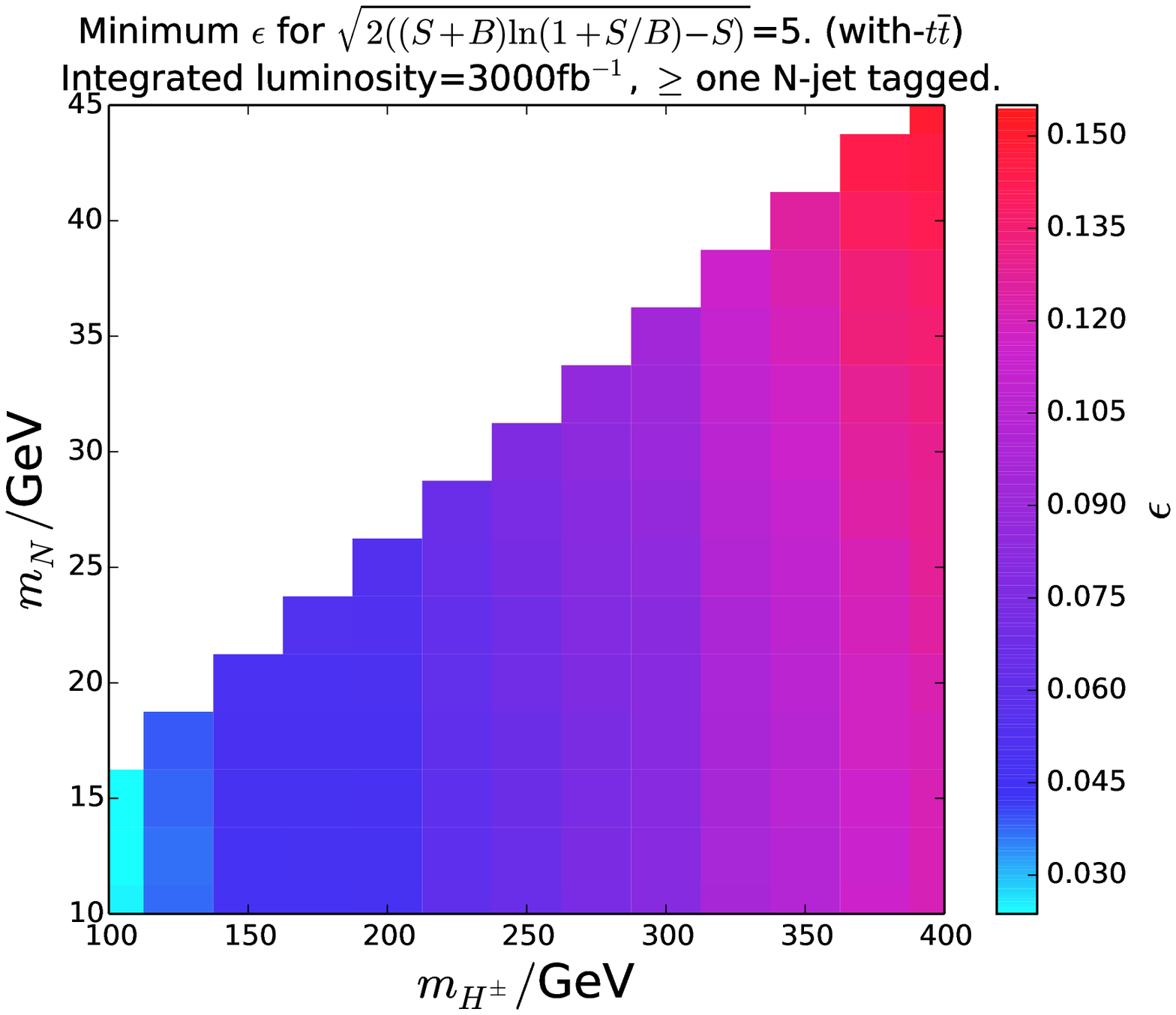}
\caption{Minimum $\epsilon$ for $\sqrt{2 ((S+B) \ln(1+S/B)-S)} = 5$. The integrated luminosity is set 3 ab$^{-1}$ on a 13 TeV proton-proton collider. $p p \rightarrow t \overline{t} \rightarrow \mu^+ \mu^- b \overline{b} \nu \overline{\nu}$ contributions are included.} \label{MinimumEpsilon_3000_tt}
\end{figure}

\begin{figure}
\includegraphics[width=3.2in]{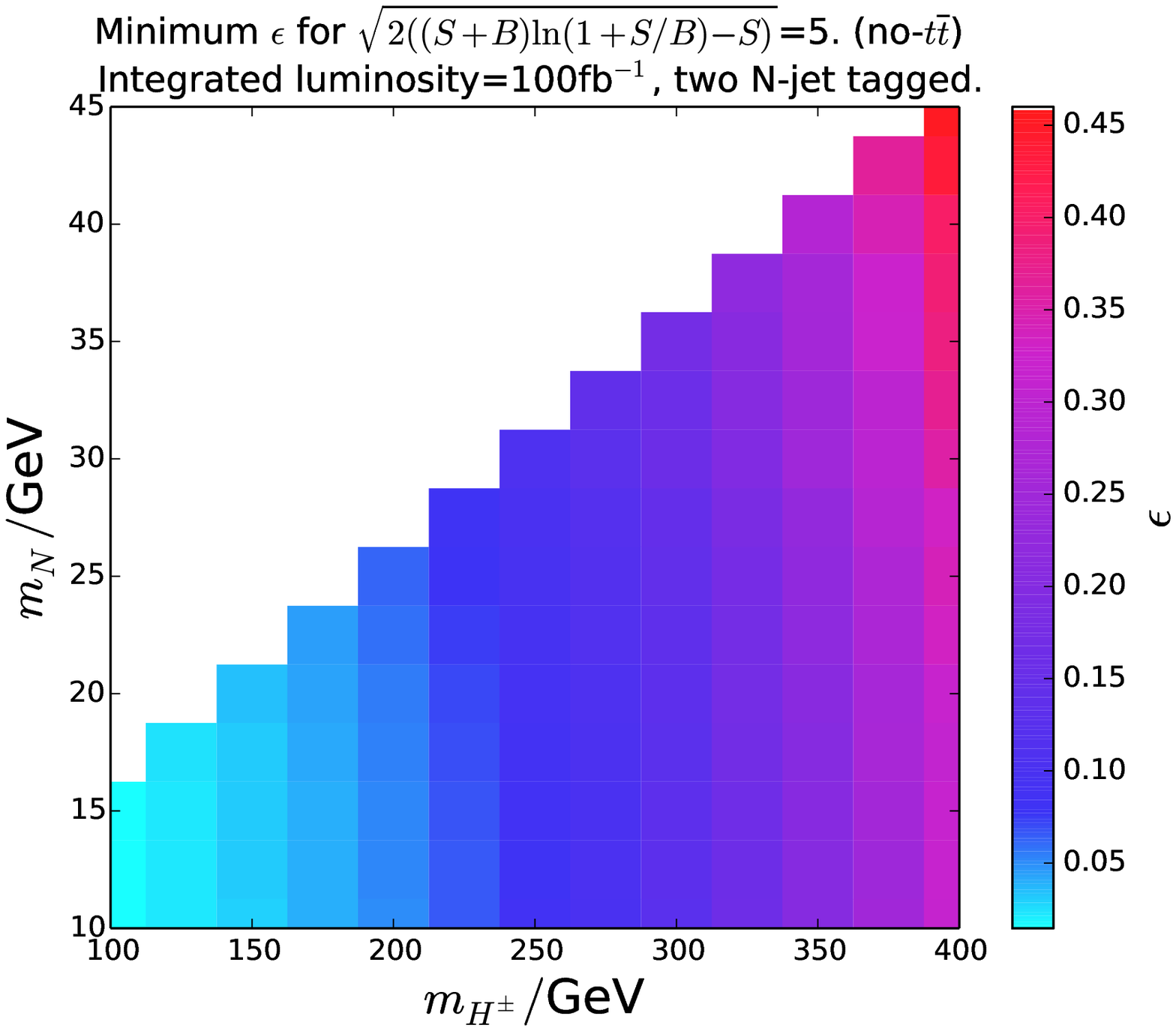}
\includegraphics[width=3.2in]{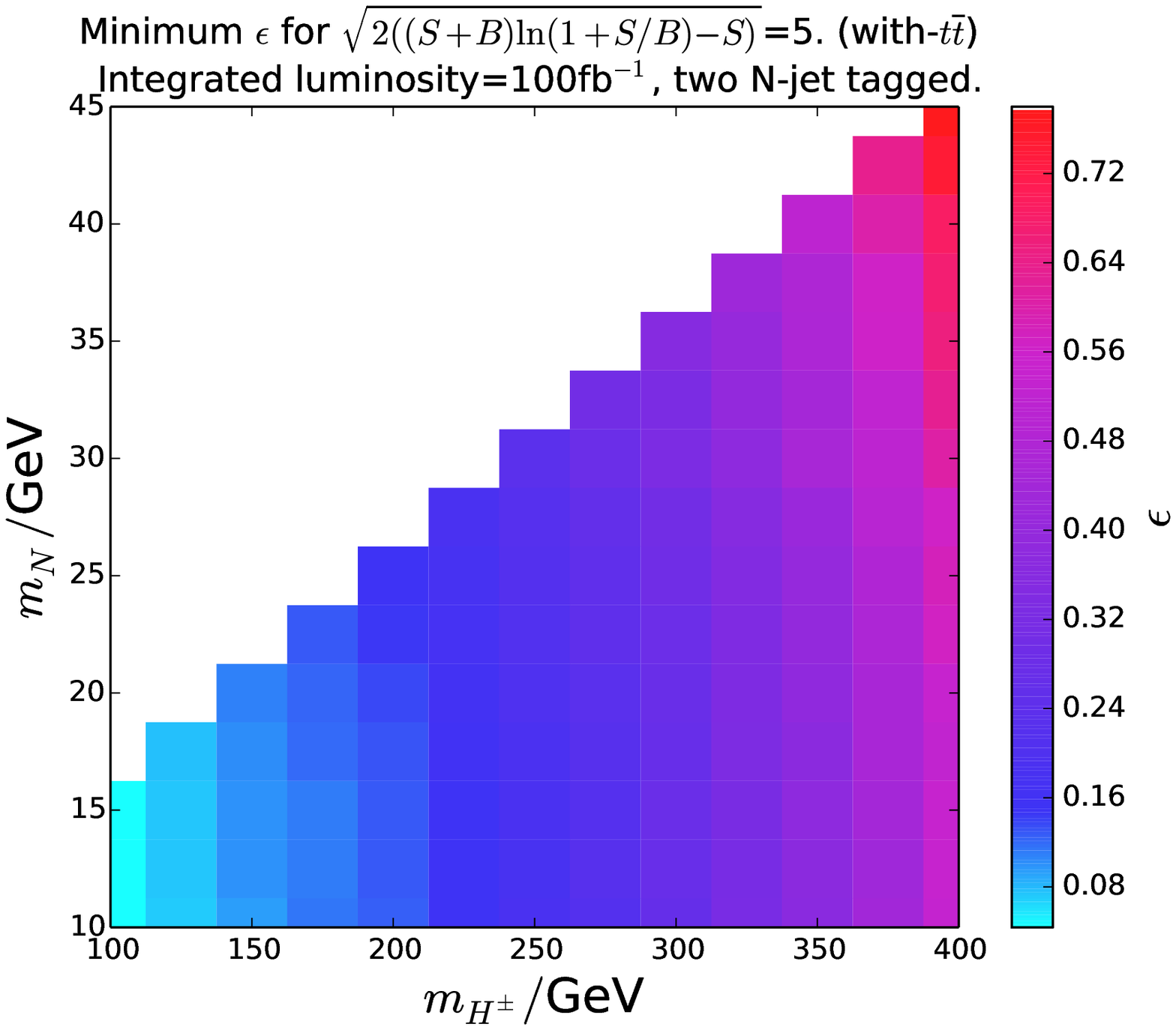}
\caption{Minimum $\epsilon$ for $\sqrt{2 ((S+B) \ln(1+S/B)-S)} = 5$. The integrated luminosity is set 100 fb$^{-1}$ on a 13 TeV proton-proton collider. In the left panel, contributions from the $p p \rightarrow t \overline{t} \rightarrow \mu^+ \mu^- b \overline{b} \nu \overline{\nu}$ are not included. While in the right panel, they are included. } \label{MinimumEpsilon_100}
\end{figure}

In the Fig.~\ref{MinimumEpsilon_3000}, \ref{MinimumEpsilon_3000_tt}, we have shown the minimum $\epsilon$ for the significance defined by $\sqrt{2 ((S+B) \ln(1+S/B)-S)} = 5$ in the luminosity of 3 ab$^{-1}$ as proposed by the HL-LHC. Both the ``-1N-jet'' and the ``-2N-jet'' results are plotted. For comparison, we have also shown in the Fig.~\ref{MinimumEpsilon_100} the results on the 100 fb$^{-1}$, which can be reached in the near future. Here, only the ``-2N-jet'' results are plotted. Again, both ``no-$t \overline{t}$'' and ``with-$t \overline{t}$'' situations are plotted. 

\section{Discussions}

In the case of the Majorana sterile neutrino, the same-sign charged lepton signals could appear. We should note that the isolated leptons decayed directly from the $H^{\pm}$ are definitely oppositely charged. Therefore we should concern the charges of the leptons appeared inside the jets. As we know, we have not seen any in the literature addressing the charge identification performance in such a case. Further more, in the pseudo-Dirac sterile neutrino case, there should not be any significant same-sign lepton signals. In order for a safe and careful discussion, and the convenience to transfer our results to the pseudo-Dirac cases, we did not discuss about such a kind of signal.

As we have mentioned, we have only calculated the results of the muon-channels, but our results can be transferred to the $H^{\pm} \rightarrow e^{\pm} N$ channels. However, it is difficult to discriminate the electron within the $N \rightarrow e q \overline{q}$ jets. If we again let the $N \rightarrow \mu q \overline{q}$ while keeping the $H^{\pm}$ still decaying to isolated electrons, the channel $H^{\pm} \rightarrow \mu^{\pm} N$ is also inevitable. Therefore a lepton flavour violation $e^{\pm} \mu^{\mp}$ signal could arise, giving a rather significant signal in the case that $t \overline{t}$ backgrounds can be effectively eliminated.

Compared with the hadron-hadron colliders, the $e^+ e^-$ collider provides a cleaner background. As we know, there are currently proposals of the Circular Electron Positron Collider (CEPC) \cite{CEPC-SPPC-1, CEPC-SPPC-2}, the International Linear Collider (ILC) \cite{ILC-1, ILC-2}, the Compact Linear Collider (CLIC) \cite{CLIC-1, CLIC-2} and the $e^+ e^-$ project of the Future Circular Collider (FCC-ee) \cite{FCC-ee-1, FCC-ee-2}. the CEPC and the FCC-ee are circular colliders with no proposals for $\sqrt{s} \gtrsim 500$ GeV as we have learnt, while both the ILC and the CLIC have plans to run at $\sqrt{s} \gtrsim 1$ TeV scale.

\begin{figure}
\includegraphics[width=2in]{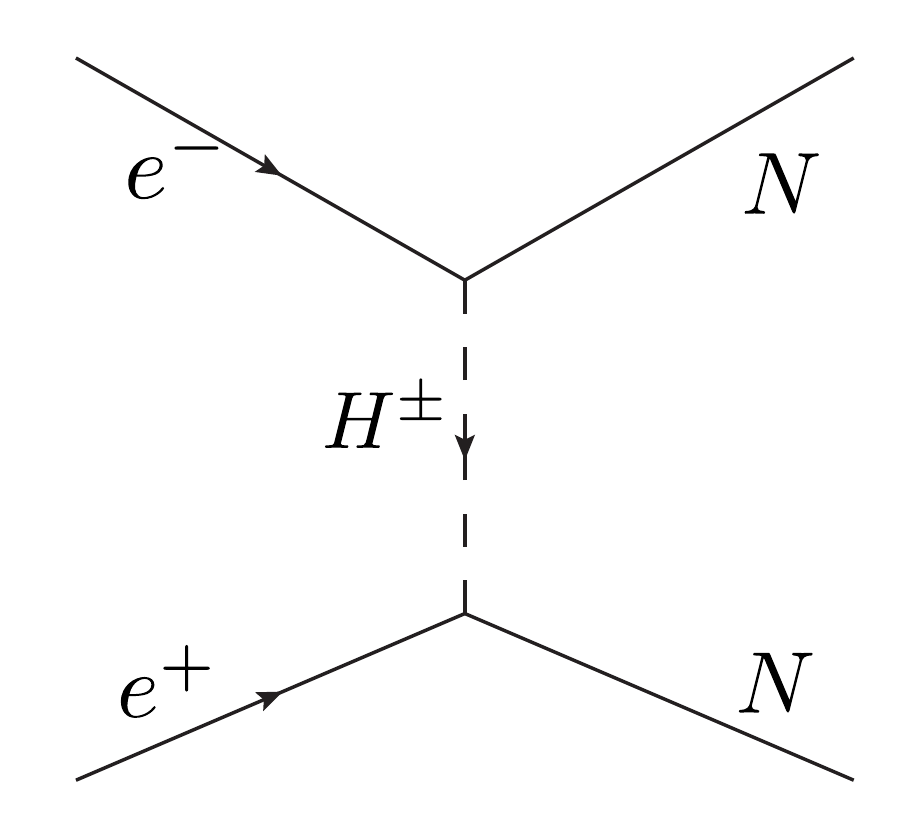}
\includegraphics[width=2in]{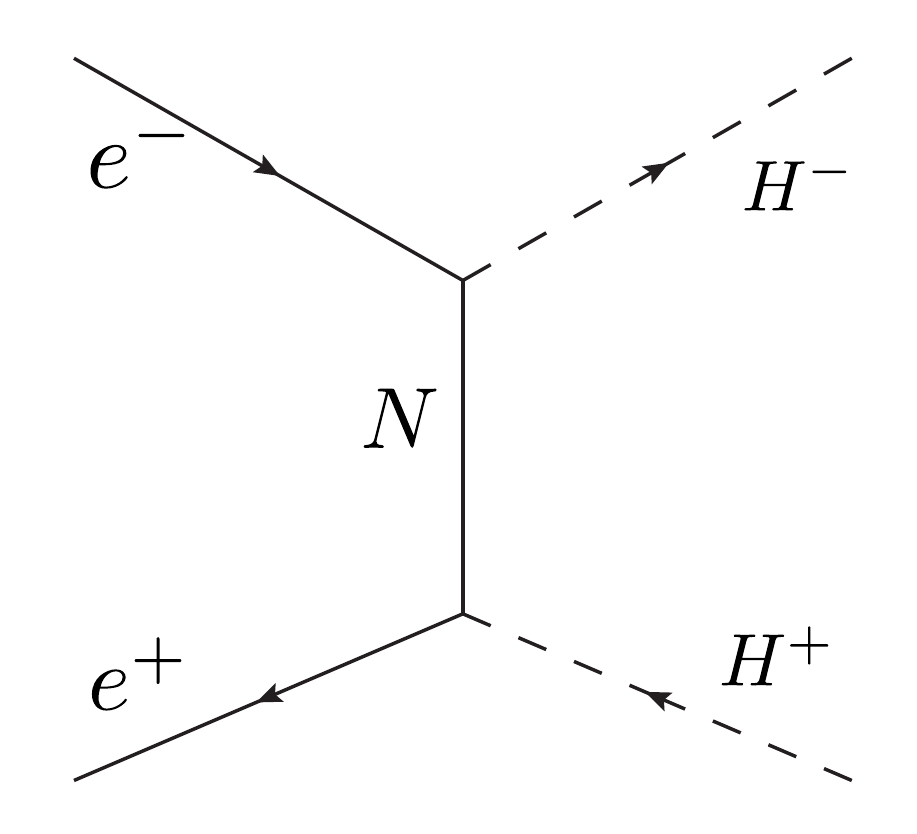}
\caption{t-channel diagrams on the $e^+ e^-$ collider.} \label{ee_t}
\end{figure}

On the $e^+ e^-$ collider, the sterile neutrino pairs can be directly produced through the t-channel diagram in the left panel of the Fig.~\ref{ee_t}. This channel might be significant due to the might-be large $e$-$H^{\pm}$-$N$ Yukawa couplings. Besides the electroweak processes similar to the Fig.~\ref{MyProcess}, charged Higgs boson can also be produced through the exchanging of a t-channel sterile neutrino in the right panel of the Fig.~\ref{ee_t}.

In the signal-background analysis on the $e^+ e^-$ collider, at least one more technique can be applied for suppressing the background. In fact, unlike the sterile neutrino, B-meson's semi-leptonic decay also produces a neutrino, therefore the missing energy arises. Due to the more precise measurement of the total missing energy and the no pile-up environment, the missing energy can also become a good kinematic handle to cut the background.

\section{Conclusions}

In this paper, we have discussed about the collider simulation in a particular parameter space in the $\nu$-THDM. The light sterile neutrinos are hereby decay product of the heavy charged Higgs boson. In the $m_N \ll m_H^{\pm}$ cases, the muons appeared in the collimated decay products of the sterile neutrinos can help us discriminate the $N$-jets from the QCD jet backgrounds. We have shown that these backgrounds can be effectively suppressed. In some parameter space, future proton-proton 3000 ab$^{-1}$ collider can be sensitive to the $\epsilon \lesssim 0.01$ cases. The reducible $p p \rightarrow t \overline{t}$ background plays an important role in the sensitivity. Further knowledges on the collider designs and performances should be well-studied before trying to eliminating this channel.

\begin{acknowledgements}
The author would like to thank for Pyungwon Ko, Taoli Cheng, Peiwen Wu and Jinmian Li for helpful discussions. This work is supported in part by National Research Foundation of Korea (NRF) Research Grant NRF- 2015R1A2A1A05001869, and the Korea Research Fellowship Program through the National Research Foundation of Korea (NRF) funded by the Ministry of Science and ICT (2017H1D3A1A01014127).

\end{acknowledgements}

\newpage
\bibliography{LightN}
\end{document}